# TCloud: A Dynamic Framework and Policies for Access Control across Multiple Domains in Cloud Computing

Sultan Ullah
School of Computer and Communication Engineering
University of Science and Technology Beijing

Zheng Xuefeng
School of Computer and Communication Engineering
University of Science and Technology Beijing

Zhou Feng
School of Computer and Communication Engineering
University of Science and Technology Beijing

## ABSTRACT
In a cloud computing environment, access control policy is an effective means of fortification cloud users and cloud resources / services against security infringements. Based on analysis of current cloud computing security characteristics, the preamble of the concept of trust, role-based access control policy, combined with the characteristics of the cloud computing environment, there are multiple security management domains, so a new cross domain framework is for access control is proposed which is based on trust. It will establish and calculate the degree of trust in the single as well as multiple domains. Role Based Access Control is used for the implementation of the access control policies in a single domain environment with the introduction of the trust concept. In multiple domains the access control will be based on the conversion of roles. On the basis of trust, and role based access control model, a new novel framework of flexible cross – domain access control framework is presented. The role assignment and conversion will take place dynamically.

## 1. INTRODUCTION
In recent years, the expansion of cloud computing has flourished in the field of computer network and information technology, and the prospects for its application are also extremely attractive. The unique advantages of cloud computing with its ultra – large – scale virtualization, high reliability lead to a great change in computing industry. However, the occurrence of security breaches shadowed the prospect cloud computing[1]. Cloud computing is the large-scale collection of network computing, virtual pool of shared resources, i.e software resources and storage resources, in order to achieve resource sharing, cloud computing must solve the problem of resource access. For every user, to interact, there is a big risk, in the case of the other identity is unknown, because the user may be a bona fide user, or it may be a malicious user. Therefore, access control strategy has become particularly important to ensure the security. Traditional access control technology need to set up a unified security management domain, a management domain within the identity-based authorization [1] [2] [3].

Cloud computing is a virtual pool of resources and provide the service over a network, and it is very obvious that each resource may not belong to the same security management domain. The network of cloud computing covers a very wide range, having cross – domain dynamic characteristics. Traditional identity-based access control technology apparently has been unable to meet the security requirements of cloud computing. Currently the most effective method is to improve and expand on the basis of the traditional access control, and thus adapt to the new demands of cloud computing security. Access control technology is a hot topic of research in cloud computing environment now a days to solve the security issues, and enable the extension of traditional access control technology [2] [3] [4] [5].

Early access control technology can not only guarantee the normal access to legitimate users, to prevent the attack of non-authorized users, and can solve the security issues caused by the operation of the legitimate user errors. In cloud computing model, the researchers are apprehensive regarding how to control access the data and resources through leading – edge ways of access control. In cloud computing access control technologies, i.e Identity and Access Management (IAM), is not very satisfactory solution in the cross – domain access control and authorization issues. Security has become an important direction of scientific research in the cross – domain access, and trust is the core of the security problem [6].

Blaze M, proposed the notion of trust management, first trust mechanism is applied to the human society, and then in technical field to provide a new way to solve the protection of data and resources problems in the cloud computing environment [7]. On the basis of trust management a trust mechanism is introduced to the access control technology. This study gives the definition and method to calculate trust and is expanded to access control entrenched in cloud computing platform. In order to solve such distributed trust issues i.e multi – domain, cloud computing environment, hierarchical management model of trust can be used. The analysis of trust model for multi – domain environment and update mechanism of the trust models are discussed, and due to the dynamic nature of trust, trusted computing implementation faces major difficulties.

In this article after analyzing the security feature of the cloud computing security, an expansion and improvement of the role based access control model based the concept of trust for multi – domain access control framework is presented. This framework presents two methods for establishing and computation of trust between local and cross – domain, and between the cloud user and cloud computing platform using dynamic authorization.

## 2. Literature Review

A Despite the new security risks and challenges presented by cloud computing as new mode of information services, but the security needs of the traditional information technology services and cloud almost remains the same. The core requirements remain to be the confidentiality of the data, applications, integrity, availability and privacy protection, and to meet these security requirement access control mechanism is the of most importance. In order to adapt to the complex access control management needs, and to achieve the dynamic demand of cloud computing environment, a role based access control model is applied and analyzed to increase the maintainability [8]. RBAC model is applicable to close centralized network environment, and is based on the identification of roles, and cannot be applied to large – scale, open distributed network to meet the needs of cross – domain environment of cloud computing. Therefore, how the cross-domain access control to authorize the urgent need to address the problem of cloud computing access control.

In addition the role assigned to the users in RBAC model only verifies the authenticity of the user's identity, without considering the credibility of the user behavior. The pre – assigned role of access authorization, regulation and control user permissions by RBAC, to assume no malicious operation will be performed, but the system is likely to have been infringed. To solve these types of problems, a number of researchers have proposed trust mechanism and is integrated into the traditional access model. The traditional RBAC model lack trust, and it is extended in Blaze trust management, based on the introduction of the concept of trust to access control mechanism propose a trust based access control model [5]. A more user centered model which is based on the user's specific requirement, consolidated variety of user trust characteristics, flexible authorization mechanism, which is safer and reasonably required permissions for the user is assigned.

A dynamic RBAC model is proposed for cloud computing environment to remove the deficiencies in RBAC model [7] [9]. The former gives the detailed calculation of the trust, and resource access control permissions assigned to information based on the user's role and trust, and to reduce the security risks of cloud computing communication and improve security. The latter are mostly theoretical analysis; solving methods did not give details of the trust. However, some of the above studies are not taking into account the presence of the characteristics of multi – tenancy, and no cross – domain access control model is developed for cloud computing.

Reliable means to provide cloud services, introduction of trust and trusted computing are considered to be hot topic in the research areas of cloud computing security and privacy. A trusted cloud computing platform is proposed, on the basis of which the cloud service provider, provides a closed box execution infrastructure which guarantees to the subscriber about the privacy and security of the running of guest virtual machine [10]. Additionally, it permits the user to launch the virtual machine instead of checking the service provider for the security and confidentiality of the service.

Sadeghi et al introduced a trusted computing technology, which not only provide trusted hardware and software but a trust mechanism as well, which is used for the confirmation of the module's behavior is trustworthy and minimize the risks in the problem of confidentiality and integrity of outsourcing data. Design a trusted software token, bound to each other with a safe and functional verification module, in order to not to disclose any information under the precondition of outsourcing encrypted sensitive data to carry out a variety of operating functions. The above study are expected to protect the security of the data resources through trusted cloud computing platform, but do not consider the trustworthiness of the cloud user [11].

Cloud computing is a multi-domain environment and usually based on security needs between security domains the firewall is used to isolate the domains and to ensure that the transfer of data between security domains should be in accordance with the access control strategy. However, existing access control models cannot meet the cross – domain access control.

## 3. Cloud Computing Trust Metrics

The human society is a complex system, and the interaction between the entities depends on the relationship of trust between each other. The traditional social model of trust mechanism form the basis for decision making for interaction between the entities with the introduction of cloud computing. The common idea of trust was first presented in the article [12], the trust in an entity is, that under certain circumstances it can be a safe, reliable way to complete the work. The article also gives the related nature of the trust relationship; the trust relationship always exists between two entities, dynamic, transitivity, specific ambiguity and uncertainty. Shortly thereafter, the trust in computer network has been widely applied to the field of P2P networks, e-commerce, grid computing, and cloud computing [13].

In real life Trust is a subjective concept, and it depends on the person's experience. The application trust to a network environment is often difficult to accurately model, or algorithm description and measurement. Trust is the identity of the entity, and it is the assessment of the reliability, integrity and performance of the entity. Based on the above concept we can classify trust into two concepts; i.e. the trust related with identity and the trust related with the behavior. The trust related with the identity is used to indicate the uniqueness of an individual, and traditional access control mechanism is often only interested to consider the identity, such as verification of identity, and entity is authorized to access appropriate resources. In realistic network environment, it cannot guarantee that an authenticated user's behavior is legal, so it is needed to trust the behavior of the entity to be taken into account.

Access control policies should be implemented on the basis of trust in the behavior. A trusted cloud computing environment can be understood as the ability of the entity to ensure safe and reliable cloud computing services. Even though this method is able to articulate trust, but it cannot determine the extent of trust. Trust is an objective reality, and in the field of access control, trust in security policy defined more clearly, and develop different security policies for different trust. Beth and Jøsang models are more typical quantitative expression of trust reasoning model [14] [15].

The Beth model introduces the concept of experience to express and assess the correlation of trust, and trust derivation formula [14]. These models of trust classify trust into two kinds which are; the direct and the indirect trust. Direct trust is the formation of trust relationship by direct interaction between entities; indirect trust is a trust relationship between the two entities which have never interacted, and established a trust relationship by the recommendation of a third party. The trust model proposed in this paper is to continue to use this classification method. The Jøsang model is based on the idea of probability theory, the concept of evidence space, and in order to express and compute the correlation of trust, this model does not explicitly distinguish among direct and recommendation trust, but provide recommendations operator for the derivation of trust [15].

The Jøsang model and Beth model cannot effectively eliminate the problem of false recommendations. The Beth model based on expansion of the proposed multi-domain access control model based on trust, and entities can own trust policy of direct trust and reliability to assign different weights for example, some entities, it gives direct trust a larger weight, and even no trust value obtained by the direct interaction of entities.

## 3.1 Trust Correlation in Single Security Domain

The traditional trusted access control model will be used for the secure operation of the system in local security domain, when entities access to other entities. In this paper, the establishment of trust relationship between cloud users and cloud resources, and the assessment of the trust of the local security domain are based on RBAC model.

Assume that $X$ is a cloud computing security management domain, the domain contains several entities $EnT \in N, while\ N > 0$, in order to calculate the degree of trust between the entities, and the implementation of local access control policy, the entities of $X$ domain needs to interact, and this interaction is based on direct or recommendation trust between entities.

**Definition 1:** The assessment of trust represented by $Ex$, and the value ranges from $-1 < Ex < 1$, and if two entities complete an interaction within the same domain, if the value of $Ex$ is positive after the interaction then it indicate the satisfaction and will increase trust, while negative value indicates not satisfied and will reduced trust. The trust value of $EnT_j$ of $EnT_i$ calculated after the completion of the $K - th$ interaction can be formalized as:

$$Ex\ (EnT_i\ , EnT_j)^k$$

**Definition 2:** in the same domain, one entity completed several interactions another, the entity's overall service satisfaction / Quality of Service will be calculated, and is denoted by $QoS$. The overall satisfaction of entity $EnT_j$ of $EnT_i$ after the completion of the $K - th$ iteration can be calculated as follows:

$$QoS\ (N_i\ , N_j)^k = \alpha \times QoS\ (EnT_i\ , EnT_j)^{k-1} + (1 - \alpha) \times Ex\ (EnT_i\ , EnT_j)^k \qquad (3 - 1)$$

**Definition 3:** An entity within the same domain, direct trust related to the assessed value, the notation used for direct trust degree is $DTD$, higher the value of domain assessment, the higher the direct trust, and the trustworthiness of the entity is higher. If there is no interaction between two entities then the value of $DTD$ is usually set to zero.

$$DTD\ (EnT_i\ , EnT_j)^k = \beta \times QoS\ (EnT_i\ , EnT_j)^{k-1} + (1 - \beta) \times Ex\ (EnT_i, EnT_j)^k \qquad (3 - 2)$$

**Definition 4:** The reputation is represented by $Rp$ and can be acquired by the entities in the same domain, if all the entities interact with the entity and satisfaction of entities in the domain. The reputation of entity $EnT_i$ in the domain $X$ can be represented by $(EnT_i, X)$ and the specific calculation formula is as follows:

$$Rp\ (EnT_i\ , X) = \frac{\sum_{j=1, j\neq i}^{k} QoS\ (EnT_i\ , EnT_j)^k \times Rp(EnT_i, X)}{k} \qquad (3 - 3)$$

**Definition 5:** The domain trust degree is represented by $TD$, and it refers to degree of trust of the entities in the domain. The $TD$ is made up of two parts, i.e. the direct trust and reputation of the entity. The trust of degree of $i - th$ entity of $EnT_i$ the $X$ domain can be formalized as $TD\ (EnT_i\ , X)$ and it can be calculated as follows:

$$TD\ (EnT_i\ , A) = \gamma \times \frac{\sum_{j=1, j\neq i}^{k} DTD\ (EnT_i\ , EnT_j)^j}{k} + (1 - \gamma) \times Rp(EnT_i\ , X) \qquad (3 - 4)$$

Wherein, $\alpha, \beta, and\ \gamma > 0$ and these weight values of the parameters are associated with the security policy of the present location - based stored in the local authentication and authorization center.

## 3.2 Trust Correlation in Multiple Security Domain

The assessment of trust degree is different for different security domains, especially for cross – domain. The influencing factor for trust on the cross domain is the behavior of the single entity in the all domains. Meanwhile, the traditional RBAC model in the cross – domain access control is no longer applicable. If the RBAC is applied to multi domain cloud environment, then the conversion of the associated role dynamically will be a problem for the model.

Suppose $X\ and\ Y$ are cloud computing environment in two different security management domains, $EnTY_i$ represent $Y$ domain entities. To calculate the degree of trust between the two entities and the implementation of cross-domain access control policy. $X$ Domain entities need to interact with the $Y$ domain entities.

**Definition 6:** Two entities of different domains interact directly which are trustworthy, the expression $DTD\ (X, Y)$ is used to calculate the value of direct trust between $X\ and\ Y$ domains. The result will continue to be updated with change in time, and it can be calculated as follows:

$$TD\ (X\ , Y) = \frac{\sum_{j=1,}^{k} DTD\ (\ EnTY_i\ , X)^j}{k} \qquad (3 - 5)$$

**Definition 7:** Entity $EnTY_i$ has a reputation in Domain $X$ and represented by $Rp\ (\ EnTY_i, X)$ and the reputation of domain $Y$ in $X$ is represented by $Rp\ (X, Y)$, and calculated as follows:

$$Rp\ (X\ , Y) = \frac{\sum_{j=1,}^{k} \theta_j \times QoS\ (\ EnTY_i\ , X)^j \times Rp(\ EnTY_j\ , X)}{k}, \sum_{j=1}^{k} \theta_j > 0 \qquad (3 - 6)$$

Where $\theta_j$ represent the weighting factor, and is relative to $EnTY_j$ in the $Y$ domain of direct trust

**Definition 8:** The inter domain trust said $TD\ (X\ , Y)$, by inter domain direct trust and reputation, both are assigned by the security management center weights:

$$TD\ (X\ , Y) = \delta \times DTD(X\ , Y) + (1 - \delta) \times Rp(X\ , Y) \qquad (3 - 7)$$

## 4. Cloud Computing Trust Metrics

The Mutual trust relationship between the user and cloud computing platform is established through the analysis of user behavior, calculated trust according to the trust model, combined with the implementation of dynamic role – based access control in cloud computing environment. The presence of multiple security management domains is the distinctiveness feature of the cloud computing environment, after the introduction of trust in the access control, a cross – domain access control model is needed to be developed for cloud computing environment, which should also be based on trust.

The main difference between the traditional and nontraditional model of access control for single and multiple security domain environments lies in the local and non local access control policies. In a cross domain access control which is based on trust, the first step is to identify the user, and then verify the user's trust identity / level of trust, according to the level of trust on the basis of which the decision is taken whether to authorize the user or declined. Similarly the model of user behavior and trust level reflects the overall trustworthiness of the user behavior. As a result authorization is no longer a simple static mechanism of identity, but now it is based on the combination of dynamic identity and behavior trust mechanism. The high level framework of multiple domain access control is depicted in the following diagram, which is based on trust.

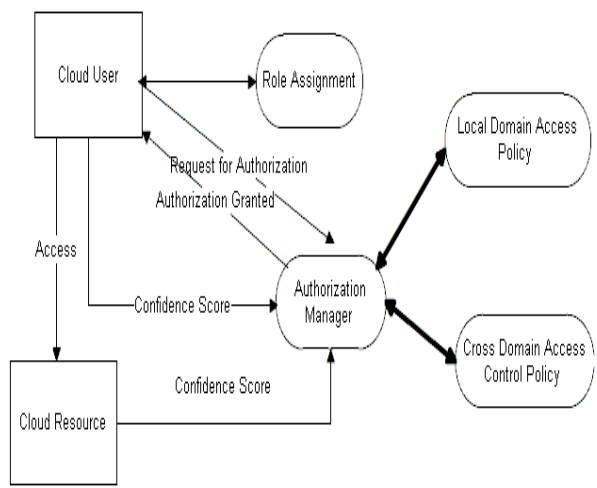

**Figure 1: Dynamic Access Control Based on Trust**

The cloud user first obtain an appropriate role by the role of the management center to submit a user ID, password, role, and want to access resources, and then interact with the authentication and authorization center to apply for authorization to access. If a user requests access to resources in the local domain, the local access control policies are in place, or to request an access outside local domain resources, cross-domain access control policies of permissions and management will be implemented.

### 4.1 Dynamic Access Control Framework for Single Domain

The introduction of trust in the role – based access control is the basic attribute of cloud users and cloud service provider for cloud resources. In single security domain trust management, authentication and authorization is performed by authentication and authorization center (AAC). . In multiple security domain trust management, authentication and authorization is performed by an advanced authentication and authorization center (AAAC).

In the local domain, every cloud users request access to cloud services or cloud resources, authentication and authorization center will see the degree of trust threshold of the cloud user, to ensure users access to cloud. The dynamic access control framework for single domain is presented in figure 2.

The steps for authentication and authorization required for single domain access control are:

i. In the role based access control, the cloud users first initiated a request for the role assignment in the access control, and thus indirectly obtain the appropriate access control permissions in this model. The role enables the users only to gain access, and the user will go through the stages of trust management in order to use these permissions.

ii. The cloud user sends access control information including user ID, Password to AAC and access resources or service ID. The AAC first authenticate the user through user's identity information and then authorize the user through the trust management based on their trust on cloud user.

The authentication Process Includes:

a. Policy Database to initialize the domain security policy.

b. Policy Enforcement Point passed user's access request to the Policy Decision Point.

c. Policy Decision Point passed the request for access information to Policy Information Point.

d. For cloud user trust and other attribute information are requested from the Trust Management Point, which in accordance with the current access Policy information and trust level an access control policy is issued.

e. Policy Decision Point, according to the information of the user trust level as well as the current security policy, to make access control decisions, and return to the Policy Enforcement Point.

f. The result is sent back by Policy Enforcement to the user entity. If the request for access of the user entity is honored, a certificate will be issued to the entity, so that users get access to the right to use its role corresponding.

iii. Cloud user to perform access control permissions, access to cloud services or cloud resources.

iv. After the interaction, users of cloud services or cloud resources performance evaluation is performed, through the trust evaluation point which calculates a new trust degree, and sent to AAC.

v. Cloud a service or resource providers also give feedback to AAC after the trust degree evaluation cloud user.

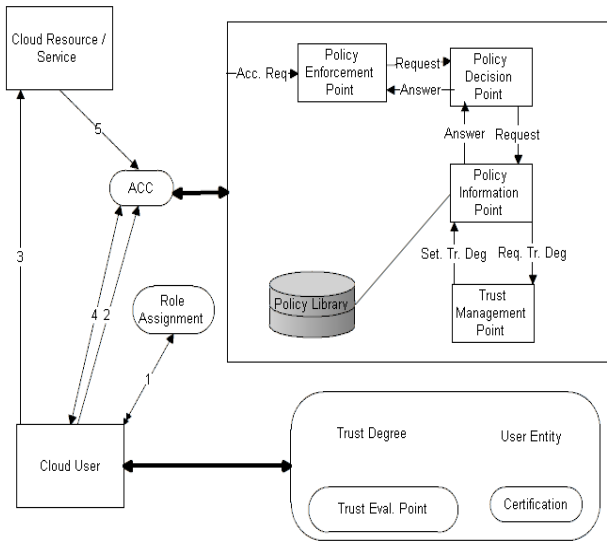

**Figure 2: Single Domain Access Control Model**

## 4.2 Dynamic Access Control for Multiple Domains

The As the cloud users often need to access a different security management domain cloud services or cloud resources, secure and effective cross domain access control is very necessary. There is not much research going on in the area of cross domain access control problem in cloud computing industry and academia, but it this problem cannot be ignored. The traditional role-based access control mechanism is applied to a closed network security environment, and therefore unable to meet the security needs of the cross – domain security environment. Therefore, the figure 3 shows the dynamic trust based access control which is the proposed framework for multiple cloud computing security domains.

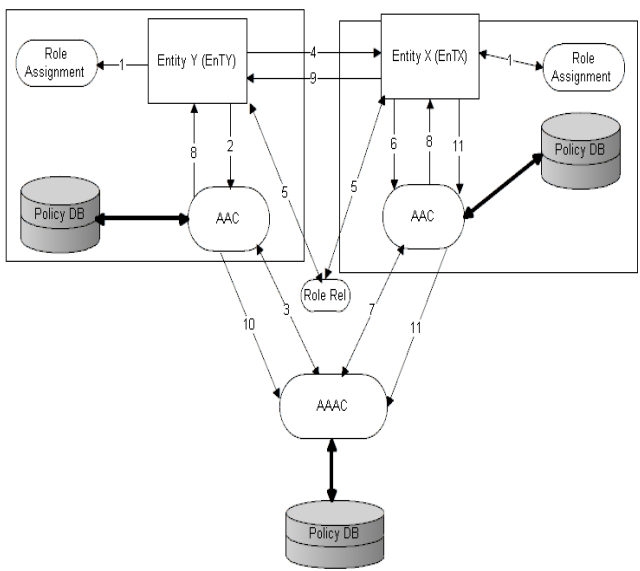

**Figure 3: Dynamic Access Control Framework for Cross Domain**

The Process of access control and role assignment on the basis of trust is as follows:

i. Entity $EnTY$ is assigned a role by role distribution center in domain $Y$, while entity $EnTX$ is assigned role by the role distribution center of $X$ domain.

ii. $EnTY$ Sends access requests to the $X$ domain $AAC$, $AAC$ calculate $EnTY$ trust, through the local security policy, AAC inform $EnTY$ domain about the decision regarding the permission of access.

iii. $ACC$ Of $X$ domain sent access request to $AAAC$, and the $AAAC$ View $X$ domain trust relationship with $Y$ domain, to determine whether access is allowed, if allowed, $AAC$ issue a certificate to $EnTY$.

iv. $EnTY$ Sent an access request containing security certificate and its role to $EnTX$ in the $X$ domain.

v. $EnTX$ Receives the request for access control of $EnTY$, first of all $EnTY$ role is converted into an associated role by $X$ domain according to the role of the $Y$ domain. It also view the permission associated with role to have the access to the resources of the $X$ domain. If this cannot be done then $EnTY$ request is directly denied.

vi. $EnTX$ Issues authentication certificate to the $X$ domain $AAC$.

vii. The $X$ domain $AAC$ contacts $AAAC$, and then performs $X$ and $Y$ domain's trust and identity mapping. Then $AAAC$ complete the link between the two domains and the transfer of the certificate. Then $AAAC$ calculates the trust of $EnTY$ in domain $X$, and the results are returned to the $X$ domain's $AAC$.

viii. $X$ Domain $AAC$ judge $EnTY's$ security attributes with the local security policy and the result is sent back to $EnTX$.

ix. The $EnTX$ will be authorized to return the result to $EnTY$, if the request is allowed, $EnTX$ allows $EnTY$ to use the resource, if denied, $EnTX$ would refuse the request of $EnTY$.

x. After completing the transaction, the $EnTY's$ resource service evaluation, and the results are sent to the $Y$ domain $AAC$, and then send to the $AAAC$.

xi. Then $EntX$ on user $EntY$ evaluation, the results are sent to the $X$ domain $AAC$, and then send to the $AAAC$.

xii. Then $AAAC$ on the basis of the evaluation value from $X, Y$ domains $ACC's$ will calculate and updates the mutual trust between the domains $X$ $and$ $Y$.

## 4.3 Inter Domain Role Conversion

Cloud computing environment consists of numerous security management domains, therefore the security and interoperability issues between these security domains needs to be considered. The first model which proposed a cross – domain access control and secures interoperability mechanism is IRBAC 2000, which achieved this by implementing the dynamic role conversion between various security domains [16].

In this article a model of access control across multiple – domain of cloud is presented, which is an extension and based on the idea of RBAC model in which the user is assigned a role, and on the basis of that role it is decided which job to execute and to which extent the user have permissions. For example if there are two security domains $X$ and $Y$, and secure interoperability is needed, then it is necessary to design a secure mechanism for the transmissions between the two domains. In order to establish a security mechanism between domains, there must some consensus on the security policy. The most basic method is the two domains can establish a default policy, i.e to provide basic security. But this does not meet the high security and reliability requirements of the multi – domain environment of cloud computing. For example, in figure 2 $EntY$ hoped that the target object $EntX$, to establish a secure environment for which the underlying security model will be the local individual security models of $X$ and $Y$ domains.

In order to obtain a higher degree of flexibility, then $EntY$ and the target object $EntX$ must know each other's identity; to cope with this situation, it is a very common problem to handle in a single domain. However, because $EntX$ and $EntY$ are in a different domain, they usually do not know each other's identity. To solve this problem, we propose a policy framework to simplify security, and interoperability between two or more domains. The framework of this policy is operated by a group of association between the local and outside domains role hierarchy. These associations constitute a combination of role hierarchy, and the role hierarchy is still in partial order.

Let the role $rNa$ in outer / non local domain convert to $rNb$ in local domain, the $rNa$ in the local domain can get access permissions for $rNb$, which is called $rNa$ associated with $rNb$, Use the symbol $rNaR_1 \rightarrow rNbR_0$, or can be directly reduced to $(rNa, rNb)$. Where $R_1, R_0$ represent all role association / relationships from $R_1$ to $R_0$, so $R_1R_0 \subseteq R_1 \times R_0$. We can define the relationship of role $rNa$ and role $rNb$: $rNa > rNb$, to indicate that $rNa$ is above $rNb$ in the role hierarchy, or "$rNa$ is an ancestor of $rNb$".

The relationship of the role is divided into two types, one is transitive role correlation, and the other is known as non – transitive correlation.

### i. Transitive Role Correlation

For the role relationship $rNaR_1 \rightarrow rNbR_0$, if for all $Xr \in R_1, XrR_1 > rNaR_1$ contains $XrR_1 > rNbR_0$, then for all $Yr \in R_0, YrR_0 < rNbR_0$. We called this relationship the transitive role relationship, and it is claimed that this relationship can be passed on.

### ii. Non - Transitive Role Correlation

Suppose we have a correlation a $rNaR_1 \rightarrow rNbR_0$, but we don't allow the ancestors of $rNaR_1$ (roles who's hierarchy is above $rNaR_1$) inherit this correlation, this is called Non-transferable Correlation, we record this as a $rNaR_1 \rightarrow NTrNbR_0$, or $(rNa, rNb) \in NT\ R_1R_0$. Role Correlation means converting outer domain role to local role that can be accepted by local domain. Once these correlations are established, all outer domain roles will be converted to local roles dynamically.

By using Transferable Correlation and Non-transferable Correlation, we can build a combination partial order in the hierarchy between local domain and outer domain role, and define a security policy. These policies can be classified into three types.

### A. Default Correlation Policy

This policy establishes minimum correlations between outer domain sets and local domain role $G_0R_1$ (Guest), the minimum role of local domain: $G_1R_1 \rightarrow G_0R_0$ for all $Xr \in R_1, if\ XrR_1 > G_1R_1$ then $XrR_1 > G_0R_0$.

### B. Clear Correlation Policy

The security administrator clearly mapped all the roles of the outer domain to the roles available in the local security domain.

### C. Partial Correlation Policy

A mapping, if not a clear policy and there exist one or more correlations other than default policy, is called Partial Clear Policy. This policy embodies real flexibility of dynamic roles conversion. In this partial order hierarchy, outer domains roles with no clear correlation could still have logical correlation through Partial Clear Policy.

In figure 4, correlation policy 1, 2 and 3 can possibly be illustrated by this. Observe role hierarchy of domain $DM0$ and $DM1$. The arrow pointing from $Xr$ to $Yr$ stands for $Xr$ is father node of $Xr$, $Xr$ is above $Yr$ in the role hierarchy. All though the structure of H0 and H1's role hierarchy is similar, they have different semantics. If a "manager" role from outer domain want to interoperate with an application in the local domain, but the application can only be accessed by local "Professor" role, in that case, the outer domain "Manager" role must be converted to meaningful local "Professor" role (correlation 1). Correlation 2 is a non-transferable correlation. We can from figure 3 the both domain has "Guest" role. If outer domain role cannot be understood by local domain, we can define a simple policy: treating all outer domains roles as local "guest" roles, namely correlation from $GuestH1$ to $GuestH0$ (correlation 3). But it's obvious that this policy lack flexibility because all outer domain roles are treated as same roles (local guest roles).

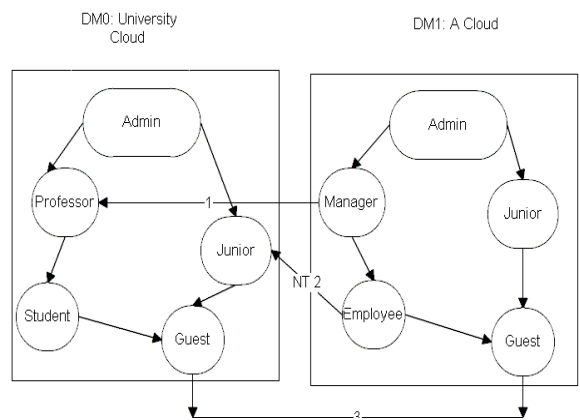

**Figure 4: Dynamic Role Conversion**

The policy frame creates a partial order relationship by adding a series of correlations between two role hierarchies. By using this mechanism, the access level of given outer domain role can be easily managed. For each outer domain role, use the highest role which is allowed by the correlations in the local role hierarchy to covert its role.

Implementation Procedure

a. The local domain security administrator establishes correlations through role editor.

b. The security server goes through all correlations and establishes a list contains all access points of local domain roles.

c. Outer domain body provide local domain policy server with its certificate.

d. Policy server adds body's access points list to the certificate of outer domain role, the conversion is accomplished.

## 5. Conclusion

In this article the access control mechanism for cloud computing environment are discussed, and proposed a dynamic model for access control across multiple domains in cloud computing based the traditional model of role based access and trust. In single domain the access is based on the traditional mechanism while in multiple security domains the roles are dynamically converted according to the domain of interest. This framework can be more intuitive, effectively protect cloud users and ensured the security of the cloud computing platform.